\documentclass[twocolumn,showpacs,showkeys,amsmath,amssymb,prl,nofootinbib]{revtex4-1}
 \usepackage{graphicx}   
\usepackage{amsmath,amssymb}        

\usepackage{color}
\definecolor{darkgreen}{rgb}{0,0.35,0}

\newcommand{\be}{\begin{equation}}
\newcommand{\ee}{\end{equation}}
\newcommand{\bea}{\begin{eqnarray}}
\newcommand{\eea}{\end{eqnarray}}

\def\6{\partial} \def\7{\tilde} \def\8{\hat}

\def\d{\dot}

\def\vs{\vskip 3mm}\def\={{\;=\;}}\def\+{{\;+\;}}

\def\ebox#1#2{\vskip 2mm{\vbox{\hrule\hbox{\vrule\kern3pt\vbox{\kern3pt
         {\begin{eqnarray}#1\label{#2}\end{eqnarray}}
         \kern3pt}\kern3pt\vrule}\hrule}}\vskip 2mm}
\def\tbox{\vskip 2mm{\vbox{\hrule\hbox{\vrule\kern3pt\vbox{\kern3pt
         {{\hfill {\small ${}^{notebook\; kiyoshi
         }$} \\
         \large \bf ~~\reptitle}\\ } 
         \kern3pt}\kern3pt\vrule}\hrule}}\vskip 2mm}

\def\vs{\vskip 4mm}

\def\dif{{\rm d}}
\def\deriv{\@ifnextchar[{\@deriv}{\@deriv[]}}
   \def\@deriv[#1]#2#3{\mathchoice%
{{\dif^{#1}#2\over\dif{#3}^{#1}}}{{\dif^{#1}#2/\dif{#3}^{#1}}}%
{{\dif^{#1}#2\over\dif{#3}^{#1}}}{{\dif^{#1}#2/\dif{#3}^{#1}}}}

\usepackage{amssymb}
\usepackage{amsmath}
\usepackage{latexsym}
\usepackage{color}

\def\d2derpar#1#2{\mathchoice%
{{\partial^2 #1\over\partial #2^2}}%
{{\partial^2 #1/\partial #2^2}}%
{{\partial^2 #1\over\partial #2^2}}%
{{\partial^2 #1/\partial #2^2}}%
}

\def\dif{{\rm d}}
\def\deriv{\@ifnextchar[{\@deriv}{\@deriv[]}}
   \def\@deriv[#1]#2#3{\mathchoice%
{{\dif^{#1}#2\over\dif{#3}^{#1}}}{{\dif^{#1}#2/\dif{#3}^{#1}}}%
{{\dif^{#1}#2\over\dif{#3}^{#1}}}{{\dif^{#1}#2/\dif{#3}^{#1}}}}

\begin{document}
\preprint{
}
\title{Canonical  Realization of BMS Symmetry. Quadratic Casimir} 
\author{Joaquim Gomis$^1$}\author{Giorgio Longhi$^2$}
\email{gomis@ecm.ub.edu}
\affiliation {$^1$Departament d'Estructura i Constituents de la Mat\`eria 
and Institut de Ci\`encies del Cosmos, 
Universitat de Barcelona, Diagonal 647, 08028 Barcelona, Spain}
\email{longhi@fi.infn.it}
\affiliation{$^2$ Dipartimento di Fisica, Universita' di Firenze, Via G.Sansone 1, 50019 Sesto Fiorentino (FI), Italy}
\date{\today}
\begin{abstract}Abstract:
We  study the canonical realization of BMS symmetry for a massive scalar field 
introduced in reference \cite{LM}. We will construct an invariant scalar product
for the generalized momenta. As a consequence we will introduce  a quadratic Casimir with
the supertranslations
\end{abstract}


\maketitle


\vs
\textit{Motivation and results}. --- 
Recently there has been a renewed interest in the BMS group \cite{BMS-1}. 
It has been shown the BMS invariance of the  gravitational scattering 
\cite{Strominger:2013jfa}, the relation among soft graviton theorems \cite{Weinberg:1965nx} and
BMS supertranslations \cite{He:2014laa}.  The relation among supertranslations,
gravitational memory and soft gravitons theorems has been also studied \cite{Strominger:2014pwa},
There is also the proposal that BMS group could be useful to understand holography in 
asymptotically 
flat space times \cite{Banks:2003vp}  \cite{deBoer:2003vf}  \cite{Arcioni:2003xx} \cite{Barnich:2010eb}.
Related to these developments has been the study of asymptotic symmetries in quantum field theories, in the case
of QED, see \cite{He:2014cra}
for the massless case and  \cite{Campiglia:2015qka} \cite{Kapec:2015ena} for the massive case. A recent overview on the whole subject is given by Strominger at Strings 2015 \cite{andystrings}.

In this work  we  study the canonical realization of the BMS symmetry for a free massive real scalar field 
 in four dimensions
introduced in \cite{LM}. The Poincare generators $P^\mu, M^{\mu\nu}$
are  written in terms of the
Fourier modes $a(\vec k) , a^*(\vec k)$ of the plane wave expansion of the Klein-Gordon field. The momentum mass-shell condition $q^2-m^2=0$ 
defines a
3d dimensional space like-hyperboloid $H_3^1$ \cite{RLN}. It is useful to introduce
a differential operator in this space $D=-m^2\Delta+3$, where $\Delta$ is the Laplace-Beltrami operator on $H_3^1$.

It happens that  the four-dimensional momenta $k^\mu$
are zero modes of D, this suggest to look for the zero modes of this operator 
in general. These are given by an infinite set of function $w_{l,m}$, 
{ defined on $H_3^1$}, { unique 
up to rescalings}, $l = 0,1,2,... \mid m\mid \leq l $ .   The explicit expression of this functions was given in \cite{LM}, see also next section. The functions $w_{l,m}$  { can be considered as} a generalization of four momenta. This allows to define the
 supertranslations 
for massive scalar field
$P_{l,m}$ in terms of $w_{l,m}$ and the Fourier modes $a(\vec k) , a^*(\vec k)$, see next section.

The transformation of the scalar field under space-time translations is  obtained from the scalar nature of the field under Poincare transformations. This not the case for supertranslations. The supertranslations acts on the field and their conjugated momenta as a non-local
linear canonical transformation.
All together $P_{l,m}, M_{\mu\nu}$ give the infinite vector representation of the BMS group.
Note that the appearance of the  BMS symmetry  introduced for the scalar 
field \cite{LM} is not  an asymptotic gauge symmetry
as in \cite{BMS-1}.

We will construct a BMS invariant scalar product
for the generalized momenta  $w_{l,m}$, or for a rescaling of them. We  write this product in  terms of an 
infinite dimensional matrix $\eta^{l,m;l',m'}$
that generalizes the  Minkowski metric $\eta^{\mu\nu}$ for the scalar product of 4d momenta $k_\mu$.
{ In a suitable} basis of these zero modes $ \eta^{l',m';l.m}=\delta^{l',l}\delta^{m',m}$.
The convergence of this scalar product has also been studied\footnote{Details and prove of the results
will presented elsewhere}.
The scalar product is the key ingredient for the definition of a configuration space, see, for example \cite{NR}, and therefore to give meaning to coordinates $x^{l.m}$ conjugated to $w_{l,m}$.

Using this infinite dimensional metric we will construct a quadratic Casimir with
the supertranslations $"P^2"=\eta^{l,m;l',m'} P_{l,m}^* P_{l',m'}$. Physically this Casimir allows us to define  the BMS masshell constraint.

 Our analysis of the scalar product and the Casimir $"P^2"$ will be useful to study 
 BMS symmetries and it  be useful to the study of scattering S-matrix, since the in and out 
states are free fields, also to study particles, strings,..  with BMS symmetries. 
The canonical realization of BMS considered in \cite{LM} and here could be extended to other fields with non-vanishing spin and to a massless fields.

\textit{Canonical Realization of BMS Symmetry}. ---

We consider a free real scalar field $\Phi(t,\vec x)$ of mass m in four dimensions. The Fourier expansion is given by
\be
\Phi (t,\vec x)=\int_{ R^3} \tilde d k[ a(\vec k) e^{-i (x\cdot k)}+a^*(\vec k)e^{i (x\cdot k)}]
\ee
where $({}\cdot {} )$ is the ususl Minkowski Lorentz invariant product and 
$\tilde dk=\frac{d^3k}{\Omega(\vec k)}, \Omega(\vec k)=(2\pi)^3 2\omega(\vec k)=(2\pi)^3 
2\sqrt{\vec k^2+m^2}$ and  $k_i=-k^i$.

The realization of the Poincare group in terms of Fourier modes of the scalar
field  $\Phi (t,\vec x)$ is given by
\be
P_{\mu}=\int_{ R^3} \tilde d k \, k_{\mu} a^*(\vec k) a(\vec k),
\ee
\be
  M^{0j}=-i 
\int_{ R^3} \tilde d k  a^*(\vec k)
 \omega(\vec k)  \frac{\partial }{\partial \vec{k}_j} a(\vec k) 
\ee
\be
M^{ij}=-i \int_{R^3} \tilde d k  a^*(\vec k) (k^i\frac{\partial}{\partial k^j}-k^j\frac{\partial}{\partial k^i})a(\vec k) 
\ee
as one check the algebra by  using the Poisson brackets  
$
\{a(\vec k), a^*(\vec k')\}=-i \Omega(\vec k)\delta^3(\vec k-\vec k')
$.

The mass-shell condition for  $\Phi(t,\vec x)$ is given
by $q^2-m^2=0$; it defines the 3d space like hyperboloid, $H_3^1$,  
in the space of momenta, with coordinates $\vec k$. The relation 
with the embedding momenta $ q_\mu$ is $q_0=\sqrt{\vec k^2+m^2},\, q_i=k_i$.  The induced metric on
$H_3^1$ is given by
\be
ds^2=\hat\eta=(\frac 1{\omega^2(\vec k)} k_ik_j-\delta_{ij})dk_i\, dk_j.
\ee

In polar coordinates 
$ds^2 = -m^2(\frac{1}{r^2+1}dr^2+r^2ds^2_{S^2})$ 
where
$r= \frac{\mid\vec{k}\mid}{m}$, and  $ds^2_{S^2}$  is the metric of the sphere. This expression, apart dimensions, is used in  \cite{Campiglia:2015qka} 
 to give a foliation of the Minkowski space-time.
 
The 3d space-like hyperboloid, $H_3^1$ has constant negative curvature \cite{RLN} $R=-\frac 1m$. It is an Euclidean AdS$_3$ space, that can be written as the coset $\frac  {SO(4,1)}{SO(3,1)}$.  The coordinates are  a global parametrization of the Euclidean AdS$_3$ space.
The Laplace Beltrami operator on $H_3^1$ is given by
\be
\Delta=\vec\nabla\cdot\vec\nabla+\frac 2{m^2}\vec k\cdot\vec\nabla+\frac 1{m^2}(\vec k\cdot\vec\nabla)^2
\ee
It is an elliptic operator and has the property $\Delta k_{\mu}=\frac 3{m^2} k_{\mu}$.
 We introduce the operator $D$ 
\be
D=-m^2\Delta +3,  
\ee
the four momenta are zero modes of D, $Dk^\mu=0$. Since we want 
to find a generalization of four momenta to construct the supertranslations this property suggest to study  all zero modes of $Df(k)=0$. 
Introducing spherical coordinates, these are given by the functions \cite{LM}
\begin{equation}
w_{l,m}(\vec{k}) = u_l(r) Y_{l,m}(\hat{k}),
\end{equation}
\begin{equation}
u_l(r) = r^l F((l-1)/2,(l+3)/2;l+3/2;-r^2),
\end{equation}
where  $F$ is the Hypergeometric function $_2F_1$,  $Y_{l,m}$ 
are the spherical harmonics and $r = \frac{\mid\vec{k}\mid}{m}$. 
{ The set $\{ w_{l,m}\}$ is not the only set of solutions of the equation 
$D f(\vec{k}) = 0$. For given values of $\{l,m\}$ there are two independent solutions of the 
equation.  The first is $\{ w_{l,m}\}$, which has a good behaviour for $r \rightarrow 0$, 
the second set of solutions is singular in $r = 0$. This is the reason for the choice 
$\{w_{l,m}\}$. }

 The functions $w_{l,m}$  span an infinite dimensional  non unitary representation of the Lorentz group. This representation  has the following properties

  i) it is the \underline{only representation}  \cite{LM}  with an invariant subspace of \underline{dimension 4}, 
 
ii)  for $l = 0, 1$  $w_{l,m}$ is the \underline{four-vector}  $k^{\mu}$ in the spherical basis

\be
w_{0,0}=\sqrt{1=r^2}Y_{0,0}=\frac{\omega(\vec k)}{m}Y_{0,0},
\ee

\be
w_{1,m}=rY_{1,m}=\frac{|\vec k|}{m}Y_{1,m}
\ee

  iii) the functions $ w_{lm}(r,\theta,\phi) $ have an asymptotic behavior like $k^{\mu}$, for all values of $l = 0,1,2,...$.

The presence of the zero modes $w_{l,m}(\vec{k}) $ enables us to define the supertranslations in terms of Fourier modes 

\be
P_{l,m}= \int _{R^3}\tilde d k\, w_{l,m}(\vec{k}) a^*(\vec k) a(\vec k).
\ee
In reference \cite{LM} it is proved that these integrals are well defined 
 and that  $M^{\mu\nu}, P_{l,m}$ verifiy the BMS algebra.

\textit{BMS Invariant Scalar product { and Quadratic Casimir}}. ---

Since we have an infinite set of function $w_{l,m}$  that generalizes the four momenta $k_\mu$, 
it is natural to ask   if there exists a scalar product, invariant under BMS, which \underline{generalizes} the usual  Minkowski metric.

The strategy to construct this scalar product will be to require 
the hermicity of Lorenz generators acting on the set 
$w_{l,m}$  or a rescaling of them.
We first consider  the boost  $K_3 = M_{0j}$. 
We look for a new basis of zero modes 
$\{k_{l,m}\}$ where the hermiticity is studied using the diagonal scalar product,
for $l> 1$,

\be
p(\vec{k},\vec{k}') = 
\sum_{l> 1} k_{l',m'}^*(\vec{k}) \eta^{l',m';l.m}  k_{lm}(\vec{k}^{\prime})
\ee
 with  
 \be
 \eta^{l',m';l.m}=\delta^{l',l}\delta^{m',m}
\ee

For $l=0,1$ we will use the ordinary Minkowski metric.

The functions $\{k_{l,m}\}$ will be obtained by a rescaling of $\{w_{l,m}\}$
\be
k_{l,m}=N(l)\frac 1{M(l)}\, w_{l,m}, \  M(l) = {{\Gamma(2)\Gamma(l+{3\over 2})}\over
{\Gamma(2+{l\over 2})\Gamma({{l+3}\over 2})}}.
\ee
The factor $M(l)$ is introduced in order to have simple behaviour for $l\rightarrow\infty$
since  we have 
$
|w_{l,m}|\leq\, \sqrt{\frac{2l+1}{4\pi}}\sqrt{1+r^2} M(l)
$.

The factor $N(l)$ is unknow it is determined by imposing the hermiticity of $K_3$. 
We have 

 \begin{eqnarray}\nonumber
K_3 k_{l,m} &=& \frac{N(l)}{N(l+1)} a_{l,m} k_{l+1,m} +
\nonumber \frac{N(l)}{N(l-1)} b_{l,m} k_{l-1,m} \\
&\equiv& \mathcal{A}_{l,m}k_{l+1,m} +  \mathcal{B}_{l,m}k_{l-1,m}.
\end{eqnarray}
where $a_{l,m} = -i (l-1)\mathcal{C}_{l+1,m}, b_{l,m} = i (l+2)\mathcal{C}_{l,m}$
and $\mathcal{C}_{l,m} = \sqrt{\frac{(l-m)(l+m)}{(2l-1)(2l+1)}}.$
%
%
%
%

The generator $K_3$ will be self adjoint if
\begin{equation}
\mathcal{B}_{l,m} = \overline{A_{l-1,m}}.
\end{equation}
which implies

\be\nonumber
[i(l+2)\mathcal{C}_{l,m}]{N(l)\over N(l-1)} =
[i(l-2)\overline{\mathcal{C}}_{l,m}]
{N(l-1)\over N(l)},
\ee

 If we define 
 $
E(l) = [ N(l)]^2,
$ we get the recurrence relation

\begin{equation}
E(l+1) = {(l-1)\over (l+3)}E(l).
\end{equation}
This recurrence  equation is meaningful only for $l \geq 2$. The solution is given by

\begin{equation}
E(l) = \frac{4!}{(l+2)(l+1)l(l-1)} E(2),\quad l \geq 2,
\end{equation}

\noindent where $E(2)$ has an arbitrary value.

The action of $K_3$ on  $\{k_{l,m}\}$, apart from a phase factor, {\color{blue} is}
\begin{equation}
K_3 k_{l,m} = \mathcal{A}_{l,m}k_{l+1,m}  + \mathcal{B}_{l,m}k_{l-1,m},
\end{equation}

where

\begin{eqnarray}\label{AB}\nonumber
\mathcal{A}_{l,m} &= -i\sqrt{(l-1)(l+3)}\mathcal{C}_{l+1,m},\\
\mathcal{B}_{l,m} &= i\sqrt{(l-2)(l+2)}\mathcal{C}_{l,m},
\end{eqnarray}

Note that

\be
\mathcal{A}_{l,m} = \overline{\mathcal{B}_{l+1,m}},\quad
\mathcal{A}_{1,m} = \mathcal{B}_{2,m}  = 0.
\ee

The action of the other generartosrs can be obtained from 
$K_1 = i [K_3, L_2],\,
K_2 =- i [K_3, L_1],  $
and the standard action of $L_3$ and $L_{\pm}$ 

\be\label{L3}
L_3 Y_{l,m} = m Y_{l,m},
\ee

\be\label{Lpm}
L_{\pm} Y_{l ,m} = \sqrt{l(l+1) - m(m \pm)}Y_{l,m}.
\ee

\noindent We obtain 
\be\label{Kpm}
K_{\pm} k_{l,m} = \pm (\mathcal{D}^{\pm}_{l,m} k_{l+1,m \pm 1} + 
\mathcal{E}^{\pm}_{l,m} k_{l-1,m \pm 1}),
\ee
where $\mathcal{A}$ and $\mathcal{B}$ are defined in equation (\ref{AB}) 
while $\mathcal{D}$ and $\mathcal{E}$ are 
\bea\nonumber
\mathcal{D}^{\pm}_{l,m} = \sqrt{l(l+1)-m(m\pm 1)}\mathcal{A}_{l,m\pm 1} - \\
\sqrt{(l+1)(l+2)-m(m\pm 1)}\mathcal{A}_{l,m}, \\
\nonumber\eea
and 
\bea\nonumber
\mathcal{E}^{\pm}_{l,m} = \sqrt{l(l+1)-m(m\pm 1)}\mathcal{B}_{l,m\pm 1} - \\
\sqrt{(l-1)(l)-m(m\pm 1)}\mathcal{B}_{l,m}, \\
\nonumber\eea

The hermicity of the Lorentz generartors implies the invariance of the scalar product under Lorentz transformations. It is also invariant under supertranslations, since these form an abelian algebra. 

If we define the supertranslations using the set of zero modes $\{k_{l,m}\}$ 

\be
P_{l,m}= \int _{R^3}\tilde d k\, k_{l,m}(\vec{k}) a^*(\vec k) a(\vec k).
\ee

The BMS algebra becomes

\bea\nonumber\label{newbms}
&[M^{i,j},  P_{l,m}] = \epsilon_{ijk} P_{l',m'} (L_k)_{l',m';l,m} \\
&[M^{0,j},  P_{l,m}] = - P_{l',m'} (K_j)_{l',m';l,m}, \\
\nonumber\eea 

\noindent where the matrices $\parallel{L_k}\parallel$ and $\parallel{K_j}\parallel$ are 
defined in equations (\ref{AB}), (\ref{L3}), (\ref{Lpm}) and (\ref{Kpm}).   This representation gives the  vector representation of the BMS group.
Let us observe that the supertranslations act as a canonical, but non local,  
transformation of the scalar field and its momentum, see \cite{LM}.

The invariant scalar product  allows to define a quadratic Casimir for the supertranslations

\be
"P^2"=\eta^{l,m;l',m'} P^*_{l,m} P_{l',m'}
\ee

One can check  $ ["P^2", BMS]=0$
using (\ref{newbms}). 

Therefore the BMS masshell condition is given by

\be
\eta^{l,m;l',m'} P^*_{l,m} P_{l',m'}=m^2_{BMS}
\ee

 This condition will be be useful to construct
 particles, strings,.. with BMS symmetries. 
 
\textit{Acknowledgments }. --- 
We acknowledge discussions with Roberto Casalbuoni and Jaume Gomis. This work has been supported in part by 
 FPA2013- 46570, 2014-SGR-104 and Consolider CPAN and  by the Spanish MINECO under project MDM-2014-0369 of ICCUB (Unidad de Ex- celencia ÔMaria de MaeztuÕ)

\end{document}